\documentclass[10pt,twocolumn]{article}
\usepackage[textwidth=7in,textheight=8.5in]{geometry}
\usepackage{graphicx}
\usepackage{amsmath} 
\usepackage{amssymb} 
\usepackage{fancyvrb} 
\usepackage{soul}
\usepackage{fancyhdr}
\usepackage{epstopdf}
\usepackage{lineno}
\makeatletter
\renewcommand{\maketitle}{\bgroup
\begin{flushleft}
  \begin{Huge}
  \textbf{\@title}\\
  \end{Huge}
  \vspace{1cm}
  \@author
\end{flushleft}\egroup
}
\makeatother
\title{Azimuthal Anisotropy in High Energy Nuclear Collision - An Approach based on Complex Network Analysis}
\author{%
    \textbf{{\large Susmita Bhaduri}}$^{1}$, \textbf{{\Large Dipak Ghosh}}$^{2}$\\
    $^{1}$Deepa Ghosh Research Foundation, Kolkata-700031,India \\
    $^{2}$Deepa Ghosh Research Foundation, Kolkata-700031,India \\
    \underline{$^{1}$susmita.sbhaduri@dgfoundation.in}\\
    \underline{$^{2}$dipak.ghosh@dgfoundation.in}
}

\begin{document}
\twocolumn[
  \begin{@twocolumnfalse}
    \maketitle
  \end{@twocolumnfalse}
  ]
\noindent
\begin{abstract}
Recently, a complex network based method of Visibility Graph has been applied to confirm the scale-freeness and presence of fractal properties in the process of multiplicity fluctuation. Analysis of data obtained from experiments on hadron-nucleus and nucleus-nucleus interactions results in values of \textit{Power-of-Scale-freeness-of-Visibility-Graph-(PSVG)} parameter extracted from the visibility graphs. Here, the relativistic nucleus-nucleus interaction data have been analysed to detect \textit{azimuthal-anisotropy} by extending the Visibility Graph method and extracting the average clustering coefficient, one of the important topological parameters, from the graph. Azimuthal-distributions corresponding to different pseudorapidity-regions around the central-pseudorapidity value are analysed utilising the parameter. Here we attempt to correlate the conventional physical significance of this coefficient with respect to complex-network systems, with some basic notions of particle production phenomenology, like clustering and correlation. Earlier methods for detecting anisotropy in azimuthal distribution, were mostly based on the analysis of statistical fluctuation. In this work, we have attempted to find deterministic information on the anisotropy in azimuthal distribution by means of precise determination of topological parameter from a complex network perspective. 
\end{abstract}
\linebreak

\section{Introduction}
\label{intro}
Many authors have probed the azimuthal anisotropy of the produced particles in ultra-relativistic heavy-ion collisions as a function of transverse momentum and it has been used as one of the major observables to study the \textit{collective properties} of nuclear matter~[Ex. \cite{Ollitrault1992}]. The initial volume enclosing the interacting nucleons is essentially anisotropic in coordinate space, because of the geometry of non-central heavy-ion collisions. The initial coordinate space anisotropy of the overlapping zone of the colliding nuclei, in which the produced nuclear matter thermalize transforms via reciprocal interactions into the final state anisotropy in the momentum space. This area has been a field of immense interest in the recent past. 

The azimuthal anisotropic distribution in momentum space has been analysed using Fourier series~\cite{Voloshin1996}, where the first few harmonicas have been referred to as directed flow, elliptical flow etc, and in general different harmonics will have different symmetry planes. In case of an idealized initial geometry of heavy-ion interactions, all symmetry planes coincide to the reaction plane of the collision, which is constituted by the impact parameter and the beam axis. If one uses just the orthogonality properties of trigonometric functions, the Fourier series can produce some non-vanishing flow harmonics which can not confirm whether the azimuthal anisotropic distribution in momentum space has originated from a collective anisotropic flow or from some other fully unrelated physical process capable of yielding event-by-event anisotropies(as for example mini-jets)~\cite{Bilandzic2014}. 
Hence, more rigorous attempts were made to analyse collective behavior from different perspectives which could disentangle it from the processes which normally involve only a smaller subset of the produced particles termed as \textit{nonflow}~\cite{Bilandzic2014}.
The use of correlation-based techniques by including two or more particles, have eventually lead to multi-particle correlation techniques. Recently, Bilandzic et. al. have suggested that if all produced particles are independently emitted and correlated only to a few common reference planes, then the presence of \textit{azimuthal anisotropy} can be confirmed~\cite{Bilandzic2014}. This has already been confirmed mathematically in~\cite{Danielewicz1983}. 
Sarkisyan~\cite{Sarkisyan2000} has analysed the parametrization of multiplicity distributions of the produced hadrons in high-energy interaction to describe higher order genuine correlations~\cite{Abbi1999}, and established the necessity of incorporating the multiparticle correlations with the property of self-similarity to achieve a good description of the measurements.
Wang et. al.~\cite{Wang1991} and Jiang et. al.~\cite{Jiang1992} were the first to go beyond two-particle azimuthal correlations, in terms of experimental analysis. However, it did not work for increased number of particles in such multiplets. The joint probability distribution of $M$ number of particles with an $M$-multiplicity event has been applied theoretically, for the first time, in flow analysis of global event shapes~\cite{Danielewicz1983} and then in other studies~\cite{Ollitrault1992}. Borghini et. al. have further reported a series of analysis on multi-particle correlations and cumulants~\cite{Borghini2001}. Two and multi-particle cumulants had drawbacks stemming from trivial and non-negligible contributions from autocorrelations, which generates interference among various harmonics. Then Lee-Yang Zero(LYZ) method~\cite{Bhalerao2003,Bhalerao2004} filters out the authentic multi-particle estimate for flow harmonics, equivalent to the asymptotic behaviour of the cumulant series. But this approach has its own integral systematic biases. Most recently, Bilandzic et. al. have proposed the $Q$-cumulants by implementing Voloshin's fundamental idea of manifesting multi-particle azimuthal correlations in terms of $Q$-vectors assessed for different harmonics~\cite{Bilandzic2011}. Though previous drawbacks are partially removed the method  is very monotonous as calculation, and hence could be accomplished only for a small subset of multi-particle azimuthal correlations. Bilandzic et. al. have provided a generic framework which allows all multi-particle azimuthal correlations to be evaluated analytically, with a fast single pass over the data~\cite{Bilandzic2014}. It removed previous limitations and new multi-particle azimuthal observables could be obtained experimentally. But in this method a systematic bias has been found, when all particles got divided into two groups - one of reference particles and the other particles of interest.

The evidences of self-similar characteristics in high-energy interactions have connections to the idea of fractality. In view of these, study of \textit{azimuthal anisotropy} can also be attempted using different methods that are based on fractality of a complex system.
It started from the introduction of intermittency by Bialas and Peschanski~\cite{bialash986}, for the the analysis of large fluctuations, where the power-law behaviour (indicating self-similarity) of the factorial moments with decreasing size of phase-space intervals was confirmed. 
A relationship between the anomalous fractal dimension and intermittency indices has been established by Paladin \textit{et al.}~\cite{paladin1987}. The evolution of scaling-law (thereby self-similarity) in small phase-space domains has been reviewed in terms of particle correlations and fluctuations in different high-energy multiparticle collisions by De Wolf \textit{et al.}~\cite{dewolf1996}, and eventually a relationship between fractality and intermittency in multiparticle final states has been 
established. The built-in cascading mechanism in the multiparticle production process~\cite{bia1as1988}, naturally gives rise to a fractal structure to form the spectrum of fractal dimensions, and hence the presence of scale invariance in the hadronization process is evident.
Further, various methods based on the fractal theory have been utilised to examine the multiparticle emission data~\cite{hwa90,paladin1987,Grass1984,hal1986,taka1994}, and two of them, the $Gq$ moment and $Tq$ moment methods were developed, respectively by Hwa and Takagi and implemented extensively to similar systems~\cite{dghosh2012}.
Then techniques like the Detrended Fluctuation Analysis (DFA) method~\cite{cpeng1994}, Multifractal-DFA (MF-DFA) method~\cite{kantel2001} were introduced for analysing fractal and multi-fractal behaviour of fluctuations in high-energy interactions.

Recently, novel approaches to analyse complex networks have been proposed. Various natural systems can be termed as complex, and heterogeneous systems consisting of various kinds of fundamental units which communicate among themselves through varied interactions (viz. long-range and short range interactions).
Complex network based systems present us with a quantitative model for large-scale natural systems (in the various fields like physics, biology and social sciences). The topological parameters extracted from these complex networks provide us with important information about the nature of the real system.
The latest advances in the field of complex networks, have been reviewed and the analytical models for random graphs, small-world and scale-free networks have been analysed in the recent past~\cite{Albert2002,Barabasi2011}. Havlin \textit{et al.} have reported the relevance of network sciences to the analysis, perception, design and repair of multi-level complex systems which are found in man-made and human social systems, in organic and inorganic matter, in various scales (from nano to macro), in natural and in anthropogenic systems~\cite{Havlin2012}. Zhao \textit{et al.} have investigated the dynamics of stock market, using correlation-based network and identified global expansion and local clustering market behaviors during crises, using the heterogeneous time scales~\cite{Zhao20161}.

Lacasa \textit{et al.} have introduced a very interesting method of Visibility Graph analysis~\cite{laca2008,laca2009} that has gained importance because of its completely different, rigorous approach to estimate fractality. They have started applying the classical method of complex network analysis to measure long-range dependence and fractality of a time series~\cite{laca2009}. Using fractional Brownian motion (fBm) and fractional Gaussian noises (fGn) series as a theoretical framework, they have experimented over real time series in various scientific fields. 
They have converted fractional Brownian motion (fBm) and fractional Gaussian noises (fGn) series into a scale-free visibility graph having degree distribution as a function of the Hurst parameter associated with the fractal dimension which is the degree of fractality of the time-series and can be  deduced from the Detrended Fluctuation Analysis (DFA) of the time-series~\cite{kantel2001}). 
Recently, multiplicity fluctuation in $\pi^{-}$-AgBr interaction at an incident energy of $350$ GeV and $^{32}$S-AgBr interaction at an incident energy of $200$A GeV have been analysed using visibility graph method~\cite{Bhaduri20167} and the fractality of void probability distribution in $^{32}$S-Ag/Br interaction at an incident energy of $200$ GeV per nucleon has also been analysed, using the same method~(see ~\cite{Bhaduri20172} and reference there in).

Motivated by the findings obtained from the previous studies in this work the \textit{azimuthal anisotropy} was studied using the $^{32}$S-AgBr interaction at $200$A GeV by extending the complex network based visibility graph method. The average clustering coefficient~\cite{Watts1998}, - one of the important topological parameters, is extracted from the visibility graph constructed from the azimuthal distribution data corresponding to several pseudorapidity regions around the central pseudorapidity. 
The scale-freeness, fractal and multi-fractal properties of the process of multiparticle production, have already been confirmed in~\cite{Albajar1992,Suleymanov2003,YXZhang2007,Ferreiro2012}, by using the DFA and MF-DFA methods. Recently, complex network based visibility graph method has been applied over data collected from $\pi^-$-AgBr interaction at $350$ GeV and $^{32}$S-AgBr interaction at $200$A GeV, and then by analysing the \textit{Power of Scale-freeness of Visibility Graph (PSVG)}~\cite{laca2008,laca2009,ahmad2012} parameter extracted from the graphs, the scale-freeness and fractal properties of the process of particle production have been established~\cite{Bhaduri20167,Bhaduri20172}. Mali \textit{et al.} have applied visibility, horizontal visibility graphs and the sandbox algorithm to analyse multiparticle emission data in high-energy nucleus-nucleus collisions in~\cite{Mali2017,Mali2018}.
The topological parameters of the visibility graphs have their usual significance with respect to the complex network systems.
Here we attempted to correlate the physical significance of average clustering coefficient with some fundamental notions of particle production phenomenology, like clustering and correlation. 
Earlier methods for detecting anisotropy in azimuthal distribution, were mostly based on the analysis of statistical fluctuation. 
So, in this work, we have attempted to analyse the azimuthal distribution using the approach of complex network which gives more deterministic information about the anisotropy in azimuthal distribution by means of precise topological parameters. 

The rest of the paper is organized as follows. The method of visibility graph algorithm and the significance of complex network parameters like scale-freeness, average clustering coefficient are presented in Section~\ref{meth}. The data description and related terminologies are elaborated in Section~\ref{data}. The details of our analysis are given in Section~\ref{exp}. The physical significance of the network parameter and its prospective correlation with the traditional concepts of \textit{azimuthal anisotropy} in heavy-ion collisions, is elaborated in Section~\ref{infer}, and the paper is concluded in Section~\ref{con}.

\section{Method of analysis}
\label{meth}
As per the visibility graph method, a graph can be formed for a time or data series according to the visibility of each node from the rest of the nodes~\cite{laca2009}. In this way the visibility graph preserves the dynamics of the fluctuation of the data present within it. Hence periodic series is transformed to a regular graph, random series to a random graph and naturally fractal series to a scale-free network in which the graph's degree distribution conforms to the power-law with respect to its degree. Thus a fractal series can be mapped into a scale-free visibility graph~\cite{laca2009}, that too from a series with a \textit{finite} number of points~\cite{jiang2013} as against the other non-stationary, nonlinear methods like DFA, MF-DFA which require a \textit{infinite} data series as input. 

\subsection{Visibility Graph Method}
\label{meth:visi}
\begin{figure*}[t]
\centerline{
\includegraphics[width=4in]{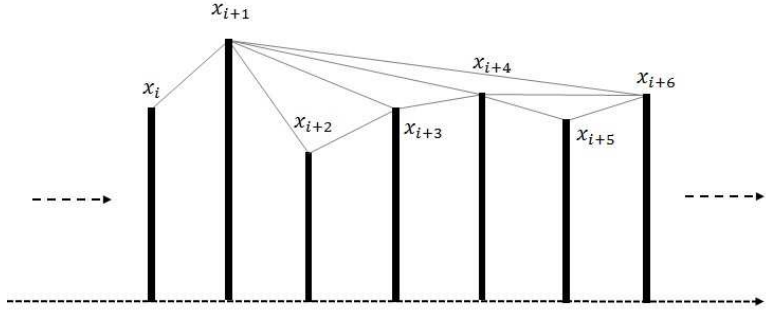}
}
\caption{Visibility Graph for time series X}
\label{visi}
\end{figure*}
Let's suppose that the value of the $i^{th}$ (in the sequence of the input data series) point of time series is $X_{i}$. In this way all the input data points are mapped to their corresponding nodes or vertices (according to their value or magnitude). In this node-series, two nodes, say $X_{m}$ and  $X_{n}$ corresponding to the $m^{th}$ and $n^{th}$ points in the time series, are said to be visible to each other or in other words joined by a two-way edge, if and only if, the following equation is satisfied. In this way, a visibility graph is constructed out of a time series $X:$ 

\setcounter{equation}{0}
\begin{equation}
X_{m+j} < X_{n} + (\frac{n - (m+j)}{n-m})\cdot(X_m - X_n) 
\label{ve}
\end{equation}
\begin{math}
\mbox{where }
\forall j \in Z^{+} \mbox{ and } j < (n-m).\\
\\
\end{math} 
The nodes $X_{m}$ and  $X_{n}$ with $m=i$ and $n=i+6$ are shown in the Figure~\ref{visi} where the nodes $X_{m}$ and $X_{n}$ are visible to each other and connected with a bi-directional edge as they satisfy Eq.~\ref{ve}. 
It is evident that the sequential nodes are always connected as two sequential points of the time or data series can always see each other.

The degree of a node of the graph \mbox{-} here visibility graph, is the number of connections/edges a node has with the rest of the nodes in the graph. Hence, the degree distribution of a network, say $P(k)$, is defined as the fraction of nodes in the network having degree $k$. Let's assume that there are $n_k$ number of nodes with degree $k$ and $n$ is the total number of nodes present in a network, then $P(k) = n_k/n$ for all probable $k$-s.

According to Lacasa \textit{et al.}~\cite{laca2008,laca2009} and Ahmadlou \textit{et al.}~\cite{ahmad2012}, the degree of scale-freeness of visibility graph corresponds to the degree of fractality and complexity of the input time or data series. The manifestation of the scale-freeness property of a visibility graph, is reflected in its degree distribution, which must obey a power-law. It means, that for a visibility graph, $P(k) \sim k^{-\lambda_p}$ is satisfied, where $\lambda_p$, a constant, is called the \textit{Power of the Scale-freeness in Visibility Graph or PSVG}. 
PSVG, thus signifies the degree of self-similarity, fractality and is therefore, a measurement of complexity of the input time series and is linearly related with the fractal dimension of the time series~\cite{laca2008,laca2009,ahmad2012}. 
Also, there is an inverse linear relationship between PSVG and the Hurst exponent of the time series\cite{laca2009}.

\subsection{Average clustering coefficient}
\label{meth:nw}
We know that by definition a cluster in a network is a set of nodes with similar features. In this experiment, we have extracted clusters of particles based on a density-based algorithm proposed by Ester \textit{et al.}~\cite{Ester1996} from the visibility networks constructed from various experimental data sets. The algorithm followed is underlined below.

For a given set of points in some space, the points that are closely packed together, or points with numerous nearby neighbours are grouped together to form clusters. Here, the density of nodes has been measured in terms of number of nodes (which has a threshold value - let's denote it by $\delta$ in our experiment) between each pair of visible nodes in the visibility graph, to form clusters. That means the closeness of the nodes to be included in a particular cluster, is measured in terms of proximity as well as in terms of the visibility with respect to each other. 
For each node (let's denote by $n_a$), among all nodes that are visible from $n_a$, (let's denote them by $\{n_{b1},n_{b2},\ldots,n_{bn}\}$) those having a count less than $\delta$, between each node $\in \{n_{b1},n_{b2},\ldots,n_{bn}\}$ and the source node $n_a$, remain in the same cluster. Naturally, these nodes are both visible from and close to the source node $n_a$. 
In this way the cluster data sets are formed for each visibility graph.
Then for each of the data sets corresponding to a particular cluster, again visibility graphs are constructed and the average clustering coefficient is extracted from each of these graphs.

Clustering coefficient is the calculation of the extent to which nodes in a graph has the tendency to cluster together. 
Average clustering coefficient has been defined by Watts and Strogatz~\cite{Watts1998} as the overall clustering coefficient of a network, which is estimated as the mean local clustering coefficient of all the nodes in the network. 
For each node present in a visibility graph, the more visible its neighbouring nodes, the more correlated and clustered these neighbours will be. In this way for each node the correlation between its neighbouring nodes are calculated and finally the average clustering coefficient of the particular visibility graph is measured.
High value of this coefficient indicates the robustness of a network.

\section{Experimental details}
\subsection{Data description}
\label{data}
The experimental data has been collected by exposing Illford G5 emulsion plates to a $^{32}$S-beam of $200$ GeV per nucleon incident energy, from CERN. A Leitz Metaloplan microscope having $10\times$ ocular lens and equipped with a semi-automatic scanning stage, was used to examine the plate. Each plate was examined by two observers to increase the accuracy in detection, counting and measurement. For the angle measurement of tracks, an oil immersion-$100\times$ objective was used.
The measuring system has $1\mu m$ resolution along $X$ and $Y$ axes, and $0.5\mu m$ resolution along $Z$ axis.

In the previous works~\cite{Bhaduri20164,Bhaduri20172} the basis for event selection is already explained.
In the context of nuclear emulsion~\cite{pow1959}, after interactions the emitted particles of different categories result in shower, grey and black tracks.
\subsection{Method of analysis}
\label{exp}
In \cite{Bhaduri20167} the pseudorapidity space for $10$ overlapping intervals around the central-pseudorapidity value (denoted by $c_r$) of the $^{32}$S-AgBr interaction, has been analysed using the visibility graph method. There, it has been established that the multiplicity fluctuation in high- energy interaction, follows scaling laws in pseudorapidity space.

In this experiment, we have considered the azimuthal angles of the shower tracks belonging to four overlapping pseudorapidity intervals centered around $c_r$, from the same interaction to analyse the fluctuation and to identify the clustering pattern from a complex network perspective and attempted to detect whether the azimuthal fluctuation is also self-similar in nature and follows scaling laws. Further the presence of \textit{azimuthal anisotropy} has been probed by extracting the clusters and calculating the average clustering coefficient, for each of the four $\phi$-data sets, in the light of conventional concept of multi-particle correlations.

\begin{figure*}[h]
\centerline{
\includegraphics[width=6in]{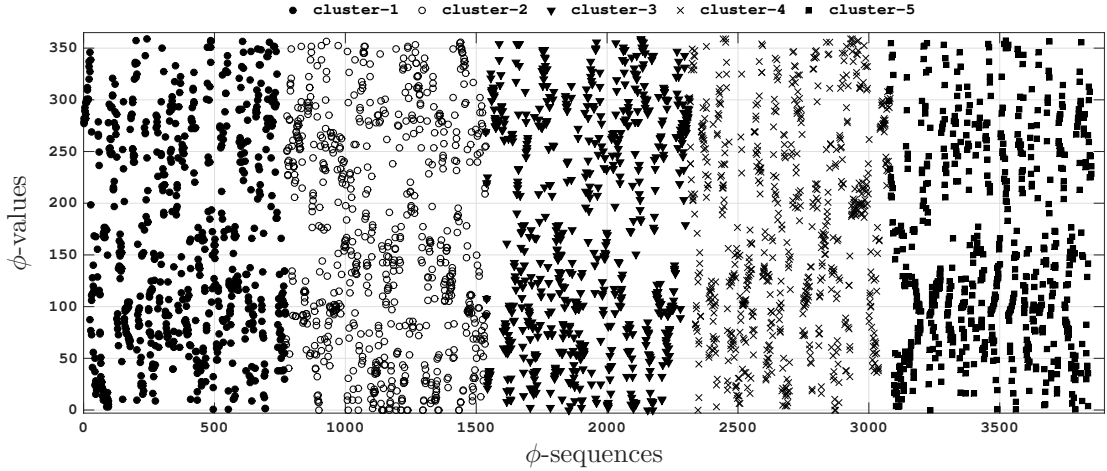}
}
\caption{$5$ clusters extracted from the azimuthal space corresponding to a sample pseudorapidity data set around $c_r$ for $^{32}$S-AgBr($200$ A GeV) interaction.}
\label{cluster}
\end{figure*}

The detailed steps for the analysis are described below.

\begin{enumerate}
\item For each of the $4$ overlapping pseudorapidity-intervals centered around $c_r$ (the range of the $\eta$-values is indicated by $c_r-\Delta\eta$ to $c_r+\Delta\eta$, where $\Delta\eta$ varies from $1$ to $4$), the $\phi$-values of the shower tracks are considered, and this way the $4$ input data sets for the experiment are formed.
Four visibility graphs are constructed according to the method described in Section.~\ref{meth:visi}. 

\item Then clusters are extracted following the density-based algorithm proposed by Ester \textit{et al.}~\cite{Ester1996} from each of the $4$ visibility graphs and one set of data points corresponding to each of the clusters, is obtained.  Figure~\ref{cluster} shows a specimen of $5$ clusters taken out from the data set of $\phi$-values in the pseudorapidity region closest to $c_r$ for $^{32}$S-AgBr(at $200$ A GeV) interaction. 

\item Once again visibility graphs are constructed from all the cluster-data extracted from the four visibility graphs constructed in the step~1. For each graph the values of $P(k)$ for all possible values of $k$-s are computed and the power-law fitting is done by following the method suggested by Clauset \textit{et al.}~\cite{Clauset2009}. 

In Fig~\ref{power_shower}(a) the $P(k)$ versus $k$ plot for a sample cluster is shown, where the power-law relationship is evident from the value of $R^2(0.96)$ of the power-law fitting. As explained in Section~\ref{meth:visi}, once the power-law has been confirmed for the $P(k)$ versus $k$ variation, the power-law exponent \textit{PSVG} of the corresponding cluster is obtained. 

\begin{figure*}[t]
\centerline{
\includegraphics[width=3.5in]{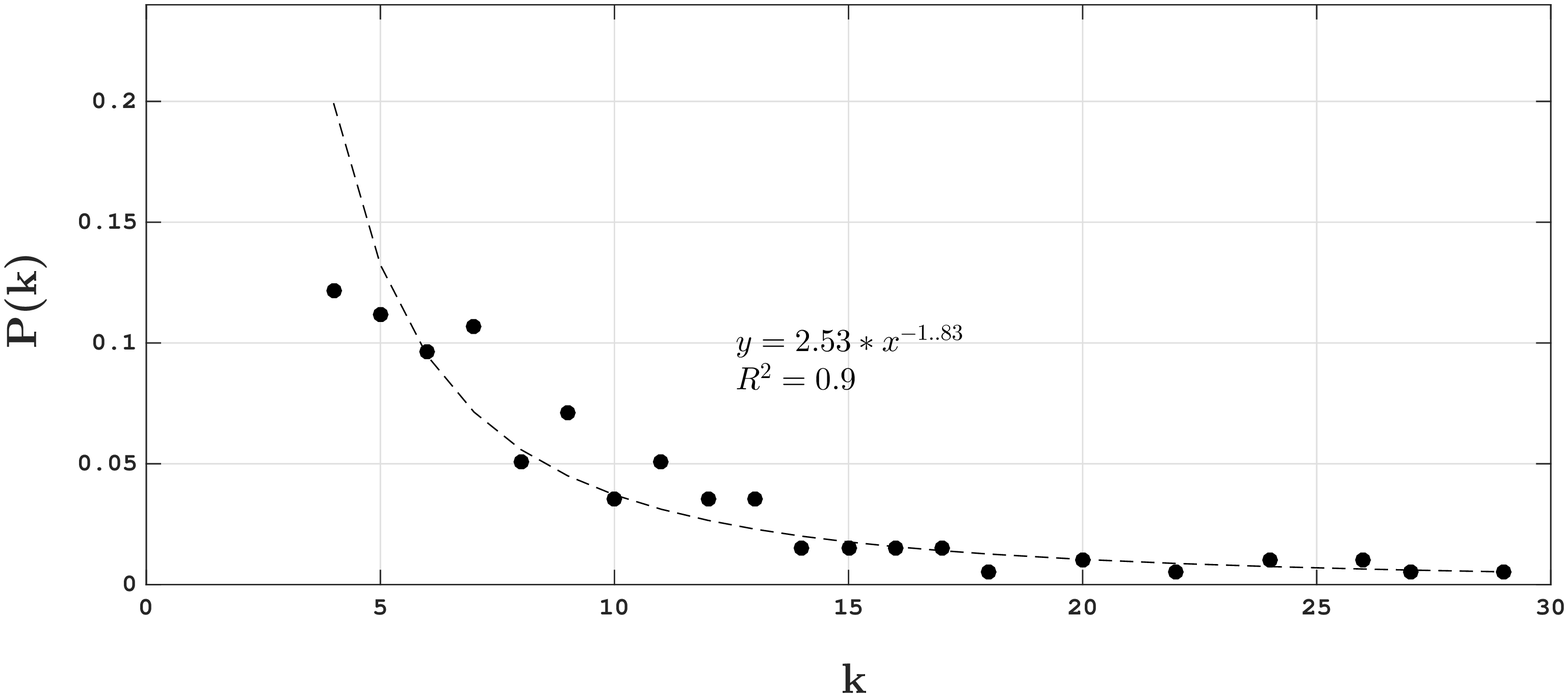}
\includegraphics[width=3.5in]{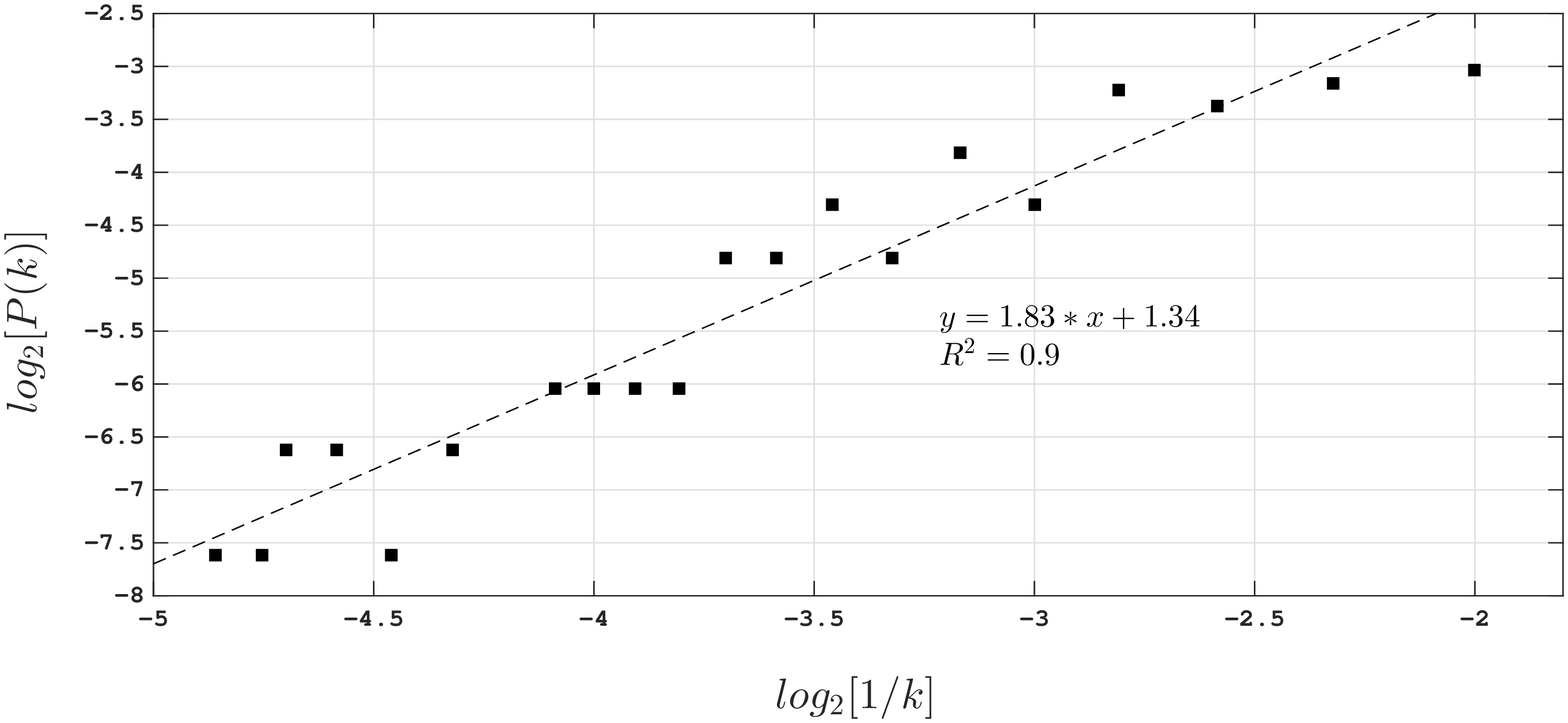}
}
\centerline{(a) \hspace*{6cm} (b)}
\caption{(a) $P(k)$ vs $k$ plot for a cluster (b) $log_{2}[P(k)]$ vs $log_{2}[1/k]$ plot for the same cluster}
\label{power_shower}
\end{figure*}

The same values of PSVG parameter are also obtained from the gradient of $log_{2}[P(k)]$ vs $log_{2}[1/k]$ plot for the same cluster data. Again goodness-of-fit for straight line fitting can be confirmed from the corresponding value of $R^2$. Fig~\ref{power_shower}(b) shows such a plot for the same specimen cluster data that is plotted in Fig~\ref{power_shower}(a) and the PSVG is 2.44$\pm$0.1, while $R^2=0.96$.

\begin{table*}[t]
\centering
\caption{List of PSVG values calculated for the visibility graphs constructed from a sample set of $30$ cluster data sets extracted from the visibility graph corresponding to one of the four data sets of $\phi$-values for $^{32}$S-AgBr ($200$ A GeV) interaction data (constructed in step~1).}
\label{cluspsvg}
\begin{tabular}{|c|c|c|c|c|c|c|c|}
\hline
\textbf{$Cluster_{seq}$}&\textbf{PSVG}&\textbf{$\chi^2$/DOF}&\textbf{$R^2$}&\textbf{$Cluster_{seq}$}&\textbf{PSVG}&\textbf{$\chi^2$/DOF}&\textbf{$R^2$}\\
\hline																																			
$	Cluster_{1}	$	&	$	2.2	\pm	0.14	$	&	$	0.92	$	&	$	0.03	$	&	$	Cluster_{16}	$	&	$	2.51	\pm	0.13	$	&	$	0.96	$	&	$	0.03	$	\\
\hline																																			
$	Cluster_{2}	$	&	$	2.03	\pm	0.18	$	&	$	0.86	$	&	$	0.03	$	&	$	Cluster_{17}	$	&	$	2.32	\pm	0.15	$	&	$	0.93	$	&	$	0.03	$	\\
\hline																																			
$	Cluster_{3}	$	&	$	2.12	\pm	0.17	$	&	$	0.88	$	&	$	0.02	$	&	$	Cluster_{18}	$	&	$	2.23	\pm	0.15	$	&	$	0.92	$	&	$	0.02	$	\\
\hline																																			
$	Cluster_{4}	$	&	$	2.1	\pm	0.14	$	&	$	0.92	$	&	$	0.02	$	&	$	Cluster_{19}	$	&	$	2.02	\pm	0.13	$	&	$	0.91	$	&	$	0.03	$	\\
\hline																																			
$	Cluster_{5}	$	&	$	2.02	\pm	0.19	$	&	$	0.85	$	&	$	0.02	$	&	$	Cluster_{20}	$	&	$	2.14	\pm	0.18	$	&	$	0.88	$	&	$	0.02	$	\\
\hline																																			
$	Cluster_{6}	$	&	$	2.15	\pm	0.18	$	&	$	0.89	$	&	$	0.05	$	&	$	Cluster_{21}	$	&	$	2.31	\pm	0.11	$	&	$	0.95	$	&	$	0.04	$	\\
\hline																																			
$	Cluster_{7}	$	&	$	2.33	\pm	0.17	$	&	$	0.92	$	&	$	0.02	$	&	$	Cluster_{22}	$	&	$	2.06	\pm	0.18	$	&	$	0.87	$	&	$	0.04	$	\\
\hline																																			
$	Cluster_{8}	$	&	$	2.25	\pm	0.2	$	&	$	0.88	$	&	$	0.02	$	&	$	Cluster_{23}	$	&	$	1.98	\pm	0.23	$	&	$	0.79	$	&	$	0.02	$	\\
\hline																																			
$	Cluster_{9}	$	&	$	2.61	\pm	0.26	$	&	$	0.86	$	&	$	0.03	$	&	$	Cluster_{24}	$	&	$	2.21	\pm	0.19	$	&	$	0.87	$	&	$	0.02	$	\\
\hline																																			
$	Cluster_{10}	$	&	$	2.3	\pm	0.13	$	&	$	0.94	$	&	$	0.02	$	&	$	Cluster_{25}	$	&	$	1.88	\pm	0.18	$	&	$	0.82	$	&	$	0.02	$	\\
\hline																																			
$	Cluster_{11}	$	&	$	2.11	\pm	0.1	$	&	$	0.95	$	&	$	0.02	$	&	$	Cluster_{26}	$	&	$	2.03	\pm	0.21	$	&	$	0.83	$	&	$	0.04	$	\\
\hline																																			
$	Cluster_{12}	$	&	$	2.16	\pm	0.14	$	&	$	0.91	$	&	$	0.03	$	&	$	Cluster_{27}	$	&	$	2	\pm	0.12	$	&	$	0.91	$	&	$	0.02	$	\\
\hline																																			
$	Cluster_{13}	$	&	$	2	\pm	0.18	$	&	$	0.87	$	&	$	0.01	$	&	$	Cluster_{28}	$	&	$	2.15	\pm	0.17	$	&	$	0.87	$	&	$	0.04	$	\\
\hline																																			
$	Cluster_{14}	$	&	$	1.83	\pm	0.16	$	&	$	0.84	$	&	$	0.04	$	&	$	Cluster_{29}	$	&	$	1.93	\pm	0.15	$	&	$	0.86	$	&	$	0.03	$	\\
\hline																																			
$	Cluster_{15}	$	&	$	2.34	\pm	0.15	$	&	$	0.93	$	&	$	0.03	$	&	$	Cluster_{30}	$	&	$	2.16	\pm	0.19	$	&	$	0.86	$	&	$	0.04	$	\\
\hline																	

\end{tabular}
\end{table*}

Table~\ref{cluspsvg} shows a list of PSVG values calculated for the visibility graphs constructed from a sample set of $30$ cluster data sets extracted from the visibility graph corresponding to one of the four data set of $\phi$-values for $^{32}$S-AgBr ($200$ A GeV) interaction data (constructed in step~1). The table also shows the corresponding $\chi^2$/DOF values, and the values of $R^2$ calculated for both power-law fitting for $P(k)$ versus $k$ data set and straight line fitting for $log_{2}[P(k)]$ vs $log_{2}[1/k]$ plot of the same data set. The power-law relationship and good scaling behaviour can be confirmed from the corresponding $\chi^2$/DOF values and the values of $R^2$ for all the cluster data sets in Table~\ref{cluspsvg}.

\item Similar analysis for all the clusters extracted from the visibility graphs constructed for all four $\phi$-data sets around the $c_r$ value of the experimental data (constructed in step~1), is carried out. As PSVG correlates the amount of complexity with fractality of the data series, and eventually with the fractal dimension of the experimental data series~\cite{laca2008,laca2009,ahmad2012}, so it can be confirmed that all these clusters are scale-free and also are of fractal structure.

\item Then Monte-Carlo simulated data set is generated for each of the cluster-data sets, assuming independent emission of pions in $^{32}$S-Ag/Br interaction at $200$A GeV. The data for Monte-Carlo simulated data sets have been chosen in such a way that $\frac{dn}{d\phi}$ distribution of the Monte-Carlo simulated data resembles the corresponding $\frac{dn}{d\phi}$ of the experimental ensembles.
Then for all these Monte-Carlo simulated data sets, first visibility graphs are constructed and then PSVG-s are calculated. Table~\ref{MCpsvg} shows the list of PSVG values along with values of $\chi^2$/DOF and $R^2$, calculated for the visibility graphs constructed from the Monte-Carlo simulated version of the same sample set of $30$ cluster data sets as shown in Table~\ref{cluspsvg} for the same particular $\phi$-data set for $^{32}$S-AgBr ($200$ A GeV) interaction (constructed in step~1) data.

\begin{table*}[t]
\centering
\caption{List of PSVG values calculated for the visibility graphs constructed from the Monte-Carlo simulated version of the same sample set of $30$ cluster data sets (as shown in Table~\ref{cluspsvg}) extracted from the visibility graph corresponding to one of the four data sets of $\phi$-values for $^{32}$S-AgBr ($200$ A GeV) interaction data (constructed in step~1).}
\label{MCpsvg}
\begin{tabular}{|c|c|c|c|c|c|c|c|}
\hline
\textbf{$Cluster_{seq}$}&\textbf{PSVG}&\textbf{$\chi^2$/DOF}&\textbf{$R^2$}&\textbf{$Cluster_{seq}$}&\textbf{PSVG}&\textbf{$\chi^2$/DOF}&\textbf{$R^2$}\\
\hline																																			
$	Cluster_{1}	$	&	$	3.24	\pm	0.11	$	&	$	0.96	$	&	$	0.02	$	&	$	Cluster_{16}	$	&	$	3.26	\pm	0.17	$	&	$	0.93	$	&	$	0.02	$	\\
\hline																																			
$	Cluster_{2}	$	&	$	2.99	\pm	0.13	$	&	$	0.94	$	&	$	0.02	$	&	$	Cluster_{17}	$	&	$	3.26	\pm	0.12	$	&	$	0.95	$	&	$	0.03	$	\\
\hline																																			
$	Cluster_{3}	$	&	$	3.29	\pm	0.13	$	&	$	0.95	$	&	$	0.02	$	&	$	Cluster_{18}	$	&	$	3.11	\pm	0.14	$	&	$	0.94	$	&	$	0.02	$	\\
\hline																																			
$	Cluster_{4}	$	&	$	3.09	\pm	0.11	$	&	$	0.96	$	&	$	0.03	$	&	$	Cluster_{19}	$	&	$	3.13	\pm	0.13	$	&	$	0.95	$	&	$	0.03	$	\\
\hline																																			
$	Cluster_{5}	$	&	$	3.14	\pm	0.14	$	&	$	0.94	$	&	$	0.02	$	&	$	Cluster_{20}	$	&	$	3.07	\pm	0.12	$	&	$	0.95	$	&	$	0.03	$	\\
\hline																																			
$	Cluster_{6}	$	&	$	2.99	\pm	0.13	$	&	$	0.94	$	&	$	0.03	$	&	$	Cluster_{21}	$	&	$	3.09	\pm	0.1	$	&	$	0.97	$	&	$	0.02	$	\\
\hline																																			
$	Cluster_{7}	$	&	$	2.97	\pm	0.1	$	&	$	0.96	$	&	$	0.02	$	&	$	Cluster_{22}	$	&	$	3.03	\pm	0.15	$	&	$	0.93	$	&	$	0.03	$	\\
\hline																																			
$	Cluster_{8}	$	&	$	2.89	\pm	0.11	$	&	$	0.95	$	&	$	0.03	$	&	$	Cluster_{23}	$	&	$	3.06	\pm	0.13	$	&	$	0.94	$	&	$	0.02	$	\\
\hline																																			
$	Cluster_{9}	$	&	$	3.18	\pm	0.17	$	&	$	0.91	$	&	$	0.02	$	&	$	Cluster_{24}	$	&	$	3.16	\pm	0.13	$	&	$	0.95	$	&	$	0.03	$	\\
\hline																																			
$	Cluster_{10}	$	&	$	3.32	\pm	0.14	$	&	$	0.94	$	&	$	0.02	$	&	$	Cluster_{25}	$	&	$	3.3	\pm	0.12	$	&	$	0.95	$	&	$	0.02	$	\\
\hline																																			
$	Cluster_{11}	$	&	$	3.22	\pm	0.13	$	&	$	0.95	$	&	$	0.01	$	&	$	Cluster_{26}	$	&	$	2.99	\pm	0.09	$	&	$	0.97	$	&	$	0.01	$	\\
\hline																																			
$	Cluster_{12}	$	&	$	3.14	\pm	0.14	$	&	$	0.94	$	&	$	0.02	$	&	$	Cluster_{27}	$	&	$	3.04	\pm	0.09	$	&	$	0.97	$	&	$	0.02	$	\\
\hline																																			
$	Cluster_{13}	$	&	$	3.02	\pm	0.09	$	&	$	0.97	$	&	$	0.02	$	&	$	Cluster_{28}	$	&	$	3.08	\pm	0.11	$	&	$	0.95	$	&	$	0.02	$	\\
\hline																																			
$	Cluster_{14}	$	&	$	3.12	\pm	0.12	$	&	$	0.95	$	&	$	0.02	$	&	$	Cluster_{29}	$	&	$	3.08	\pm	0.12	$	&	$	0.95	$	&	$	0.02	$	\\
\hline																																			
$	Cluster_{15}	$	&	$	3.07	\pm	0.13	$	&	$	0.94	$	&	$	0.02	$	&	$	Cluster_{30}	$	&	$	3.01	\pm	0.1	$	&	$	0.96	$	&	$	0.02	$	\\
\hline																	

\end{tabular}
\end{table*}

Figure~\ref{psvg_mc} also shows the trend of PSVG values calculated for the same specimen set of $30$ cluster-datasets (both experimental and the MC simulated results) as shown in Table~\ref{cluspsvg} and Table~\ref{MCpsvg}.
The PSVG-values calculated for the experimental data sets significantly differ from their Monte-Carlo simulated counterpart. This finding establishes that the degree of complexity for any cluster is not the result of the Monte-Carlo simulated fluctuation pattern, but of some dynamics present in the cluster data sets.

\begin{figure*}[h]
\centerline{
\includegraphics[width=6in]{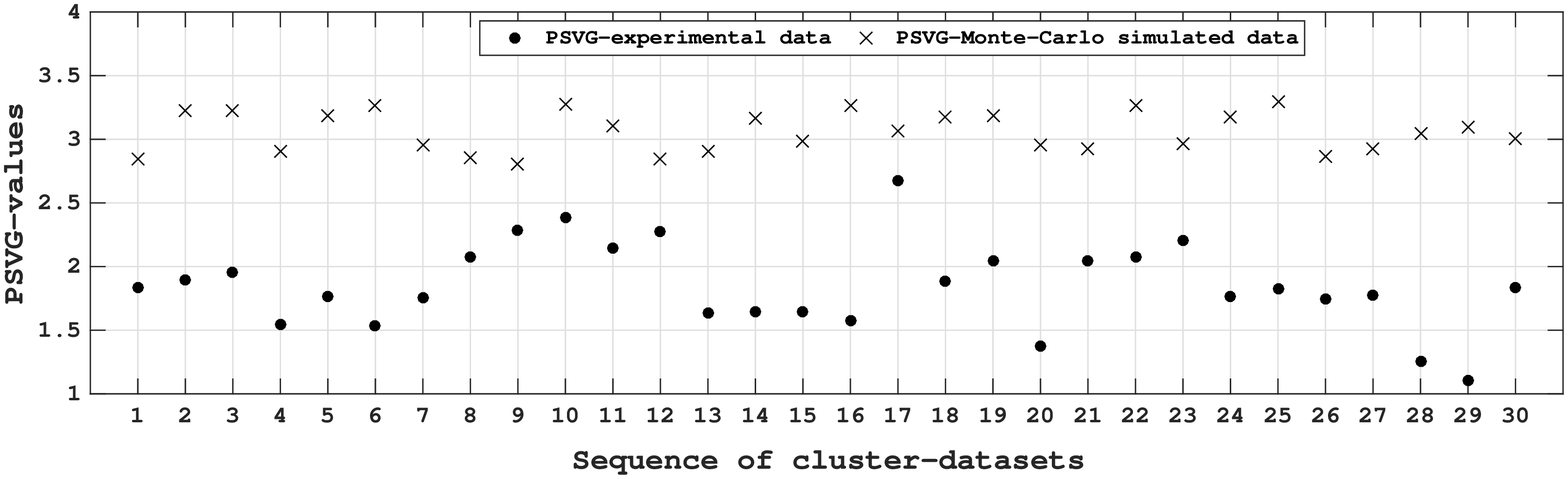}
}
\caption{Trend of PSVG values calculated for a specimen set of $30$ cluster data sets (both experimental and their MC simulated ones as shown in Table~\ref{cluspsvg} and Table~\ref{MCpsvg}) extracted from the visibility graph corresponding to one of the four data set of $\phi$-values for $^{32}$S-AgBr ($200$ A GeV) interaction data (constructed in step~1).}
\label{psvg_mc}
\end{figure*}

\item The statistical errors are shown in Table~\ref{cluspsvg}, Table~\ref{MCpsvg} and Figure~\ref{psvg_mc} obtained from the corresponding fit processes. The details of error calculation have already been discussed in a previous work~\cite{Ghosh2012}. The values of $\chi^2$/DOF and $R^2$ for the straight line fitting (for $log_{2}[1/k]$ versus $log_{2}[P(k)]$ plots), and power-law fitting(for $k$ vs $P(k)$ plots) in these figures and tables reflect the goodness of fit. We have followed Pearson's chi-squared test~\cite{Pearson1900} to calculate the $\chi^2$/DOF values, after considering the statistical errors. These errors are calculated individually for every point and the diagonal elements of the full covariance matrix are obtained thereof~\cite{Plackett1983}. The off-diagonal elements of the covariance matrix are produced because of the correlation between the data points and this correlation must be considered to have a detailed picture of the full covariance matrix. The diagonal elements of the full covariance matrix mainly contributes to $\chi^2$/DOF values, as per the analysis done in various other works~\cite{Agababyan1996}. If the off-diagonal elements are considered, the changes contributing to the $\chi^2$/DOF values are trivial and hence do not affect the final conclusion.

\item In the next step the average clustering coefficients are extracted from the visibility graphs for all the cluster data sets corresponding to the four $\phi$-data sets (constructed in step~1), as per the method explained in Section~\ref{meth:nw}. Then from the visibility graphs constructed from all the corresponding Monte-Carlo simulated data sets, the average clustering coefficients are calculated.

\begin{itemize}
\item In this way we obtain $4$ sets containing pairs of average clustering coefficients (calculated for experimental and MC simulated cluster-datasets) for the four $\phi$-data sets.
\item Each pair of average clustering coefficients corresponds to the average clustering coefficient of the visibility graph created for a particular cluster data set, and the average clustering coefficient of the visibility graph constructed also for its Monte-Carlo simulated version.
\item Then from each these four sets, the cluster-data sets which have almost similar average clustering coefficient as their Monte-Carlo simulated counterparts, are selected.
\item The count of such clusters for each of the four $\phi$-data sets, is calculated and listed in Table~\ref{clus_list}.
The trend of this count across the $\phi$-data sets (corresponding to the pseudorapidity-regions with $\Delta\eta=1.0$ to $\Delta\eta=4.0$ around $c_r$) is shown in Table~\ref{clus_list}. The trend is evidently \textit{decreasing} from the azimuthal distribution in the pseudorapidity region closest to the $c_r$ to the farthest one.
\end{itemize}
 
%

\begin{table}[h]
\centering
\caption{Trend of number of clusters are selected from each of the four $\phi$-data sets [corresponding to the pseudorapidity regions around $c_r$ for $^{32}$S-AgBr ($200$ A GeV) interaction] which have almost similar average clustering coefficient to their Monte-Carlo simulated counterparts.}
\footnotesize
\label{clus_list}
\begin{tabular}{cc}
\hline
\textbf{$\Delta\eta$}&\textbf{$Number of clusters$}\\
\hline
$1.0$&$7$\\
$2.0$&$5$\\
$3.0$&$5$\\
$4.0$&$2$\\
\hline
\end{tabular}
\end{table}
\end{enumerate}

\subsection{Results and Discussions} 
\label{infer}

As already mentioned in the section~\ref{intro}, several analysis have shown that the power-law spectra arising even from non-extensive statistics can rightly identify the general properties of particle production in high-energy interactions. Recently, Deppman et. al~\cite{Deppman2017} have shown that the fractal dimension of the thermofractal obtained from values of temperature and \textbf{entropic} index resulting from the analysis of $p_T$ distributions in high-energy interactions are similar to those found in analyses of intermittency in experimental data in high-energy collisions.

Experimental data from $^{32}$S-AgBr at $200$ A GeV interaction has been analysed using the complex-network-based visibility graph method and the PSVG values obtained thereof have been compared for all the overlapping $\eta$-regions centered around $c_r$, and it has been shown that the multiplicity fluctuation in high-energy interaction is self-similar and scale-free~\cite{Bhaduri20167}. The present work provides another evidence of fractal structure in multiparticle emission data.

In this work we have constructed visibility graphs from their corresponding azimuthal space or $\phi$-data sets. From each visibility graph, a number of cluster data sets are extracted, and then again visibility graphs are constructed for each cluster data set and its Monte-Carlo simulated counterpart. Finally it's shown that each of these clusters is self-similar, scale-free and hence is of fractal structure.

\begin{itemize}
\item For each cluster and its Monte-Carlo simulated counterpart, average clustering coefficients are calculated, and it has been found that the count of clusters having almost similar average clustering coefficient to their Monte-Carlo simulated counterparts, is decreasing from the $\phi$-region closest to the farthest from $c_r$. 
Clustering coefficient is essentially generalised as signed correlation between the nodes of the networks~\cite{Costantini2014}. In this experiment the correlation is measured in terms of the visibility between the nodes of the cluster-data sets. 

\item Hence this decreasing count of clusters shown in Table~\ref{clus_list} from pseudorapidity regions with $\Delta\eta=1.0$ to $\Delta\eta=4.0$ around $c_r$, signifies that the particle-to-particle correlation in the azimuthal distribution is \textit{least} in the pseudorapidity region that is closest to $c_r$ (with $\Delta\eta=1.0$). Therefore, this region has the maximum number of clusters having similar average clustering coefficient to their Monte-Carlo simulated versions. However, the particle-to-particle correlation is gradually increasing as the count of clusters having similar average clustering coefficient compared to their Monte-Carlo simulated counterparts decreases in the regions that are farther from $c_r$ and finally this count is least in the farthest pseudorapidity region from $c_r$.

\item Particle-to-particle correlation in the azimuthal distribution is \textit{least} in the most central pseudorapidity space, which gradually increases towards the region farthest from $c_r$ and becomes \textit{highest} in the pseudorapidity region with $\Delta\eta=4.0$.

\end{itemize}

\section{Conclusion} 
\label{con}
Bilandzic et. al. have discussed that azimuthal anisotropic distribution might be the underlying cause of collective anisotropic flow of the produced particles in ultra-relativistic heavy-ion collisions~\cite{Bilandzic2014}.
The dynamics of \textit{azimuthal anisotropy} in multi-particle production process is yet to come out with accurate parameters. Detailed and latest methods to analyse collective anisotropic flow using multi-particle azimuthal correlations, are not free from systematic bias in conventional differential flow analyses. Hence, this is still an open area of research. In view of this, we have attempted to analyse \textit{azimuthal anisotropy} by analysing the azimuthal distribution from a complex network perspective, which gives more deterministic information about the anisotropy in azimuthal distribution in terms of precise topological parameter.

It is observed that particle-to-particle correlation in the azimuthal distribution is least in the region closest to the central pseudorapidity, because in this region the pattern of clusters formed by the particles is mostly similar to their Monte-Carlo simulated counterparts. But, this correlation increases monotonically towards the region farthest from the central pseodorapidity. This in effect establishes the anisotropic nature of the azimuthal distribution closest to the central pseudorapidity region. This interesting observation might have a far reaching consequence to confirm the \textit{collective anisotropic flow} in that region.


\section{Acknowledgement} 
\label{ack}
We thank the \textbf{Department of Higher Education, Govt. of West Bengal, India} for financial support.
\section{References}

\end{document}